\documentclass[conference]{IEEEtran}
\usepackage{fancyhdr}
\usepackage{graphics}
\usepackage{listings}
\usepackage{xspace}
\usepackage{amsmath}
\usepackage{multirow}
\usepackage{rotating}
\usepackage{hhline}
\usepackage{blindtext}
\usepackage[noend]{algpseudocode}
\usepackage{etoolbox}%
\usepackage{enumitem}
\usepackage{soul}
\usepackage[colorinlistoftodos]{todonotes}
\usepackage{xcolor}
\usepackage[most]{tcolorbox}
\usepackage{setspace}
\usepackage{times}
\usepackage[T1]{fontenc}
\usepackage{amsmath}

\usepackage{cite}
\usepackage{graphicx}

\usepackage{booktabs} 
\usepackage{subcaption}
\usepackage{hyperref}
\usepackage{multirow}
\usepackage{comment}
\usepackage{setspace}

\usepackage{blindtext}
\usepackage[absolute]{textpos}

\usepackage[linesnumbered,boxed,noend]{algorithm2e}

\makeatletter
\newcommand*{\rom}[1]{\expandafter\@slowromancap\romannumeral #1@}
\makeatother

\usepackage{xr}
\definecolor{mygreen}{rgb}{0,0.6,0}
\usepackage[normalem]{ulem}

\usepackage{tikz}

\ifCLASSINFOpdf
\else
\fi
\begin{document}

\begin{textblock*}{\textwidth}(2cm,0.4cm)
Appears in the proceedings of the 33th International Parallel and Distributed Processing Symposium (IPDPS), 2019\\ \line(1,0){500}
\end{textblock*}

\title{
Runtime Concurrency Control and Operation Scheduling for High Performance Neural Network Training
}

\author{\IEEEauthorblockN{Jiawen Liu\IEEEauthorrefmark{2}, Dong Li\IEEEauthorrefmark{2}, Gokcen Kestor\IEEEauthorrefmark{3} and Jeffrey Vetter\IEEEauthorrefmark{1}}
\IEEEauthorblockA{\IEEEauthorrefmark{2}University of California, Merced\\ \IEEEauthorrefmark{3}Pacific Northwest National Laboratory\\ \IEEEauthorrefmark{1}Oak Ridge National Laboratory\\\{jliu265, dli35\}@ucmerced.edu, gokcen.kestor@pnnl.gov, vetter@ornl.gov}}

\maketitle

\pagestyle{plain}


\begin{abstract}
Training neural network (NN) often uses a machine learning framework such as TensorFlow and Caffe2. These frameworks employ a dataflow model where the NN training is modeled as a directed graph composed of a set of nodes. 
Operations in NN training are typically implemented by the frameworks as primitives and represented as nodes in the dataflow graph. 
Training NN models in a dataflow-based machine learning framework involves a large number of fine-grained operations whcih present diverse memory access patterns and computation intensity. 
Managing and scheduling those operations is challenging, because we have to decide the number of threads to run each operation (concurrency control) and schedule those operations for good hardware utilization and system throughput.

In this paper, we extend an existing runtime system (the TensorFlow
runtime) to enable automatic concurrency control and scheduling of operations. 
We explore performance modeling to predict the performance of operations with various thread-level parallelism. Our performance
model is highly accurate and lightweight. Leveraging the performance model, our runtime system employs a set of scheduling strategies that co-run operations to improve hardware utilization and system throughput. 
Our runtime system demonstrates a significant performance benefit. Comparing with using the recommended configurations for concurrency control and operation scheduling in TensorFlow, our approach achieves 36\% performance (execution time) improvement on average (up to 49\%) for 
four
neural network models, and achieves high performance close to the optimal one manually obtained by the user.
\end{abstract}


%
\IEEEpeerreviewmaketitle

\thispagestyle{plain}
\section{Introduction}
\label{sec:intro}
Neural networks (NN) have been adopted by a wide range of application domains. NN models employ increasingly larger number of parameters and data sets. Training such complex models demands immense computation and memory resources and time. Training NN models have been becoming a killer application in large-scale data centers.

Training NN models often use a machine learning (ML) framework, such as TensorFlow~\cite{tensorflow2015-whitepaper}, Caffe2~\cite{jia2014caffe} and MXNet~\cite{chen2015mxnet}. These frameworks employ a dataflow model where the NN training is modeled as a directed graph composed of a set of nodes. 
Operations, such as array concatenation, matrix multiplication and 2D convolution, are typically implemented by the frameworks as primitives and represented as nodes in the dataflow graph.
In the dataflow graph, edges between nodes capture dependencies between nodes. An operation is ready to run, as long as its dependencies (data or control dependencies) are resolved. 
 


Training NN models in an ML framework such as TensorFlow involves a large number of fine-grained operations~\cite{liu2018pim}. For example, our profiling results on training a common NN model (Inception\_v3) using TensorFlow reveal that training this NN model easily includes 16,000 operations in a single training step. Those operations have diverse memory access patterns and computation intensity. 
How to manage and schedule the operations is challenging due to the following reasons.



First, running those operations often needs to decide appropriate thread-level parallelism.
For each operation, we must choose an appropriate number of threads to parallelize the operation. In other words, we must decide \textit{intra-op parallelism} for each operation. 
Some operations do not have good scalability, because of caching effects and thread spawning overhead. Using the largest number of threads on a manycore machine to run those operations does not necessarily result in the best performance~\cite{cramer2012openmp}.
The problem of intra-op parallelism is coupled with the thread affinity problem (i.e., deciding the binding between threads and cores)~\cite{curtis2008prediction}, which makes this concurrency control problem even more challenging. 


Second, we must decide how to co-run operations. 
When operations do not have unresolved dependency and each individual operation does not sufficiently utilize hardware resources (e.g., physical cores), co-running operations may improve hardware utilization and increase system throughout.
Co-running operations decide which operations should co-run and the execution order of those operations.
In other words, we must decide \textit{inter-op parallelism} for better performance.
In essence, co-running operations is a scheduling problem.


Decisions on intra-op parallelism and intra-op parallelism for operations are often coupled. Given a manycore machine with tens of cores, we have a large search space to make the decisions.
Currently, there is no a systematic approach to efficiently control operations concurrency and schedule those large amount of operations in a NN training workload. 
The existing runtime system in machine learning frameworks simply 
uses the same intra-op parallelism for all operations and schedule operations simply according to operation dependency~\cite{performance_guide}. 
Although the ML frameworks give the user flexibility to set intra-op and inter-op parallelisms for operations, manually deciding them for all operations is impractical and often leads to poor performance~\cite{intel_optimization}.

We envision that optimizing concurrency and scheduling those operations will become more challenging, because the future NN models could involve more diverse and larger number of operations. This trend is driven by the necessity of using deeper and more complex models to improve model accuracy and the necessity of enabling dynamic interaction with execution environment to improve model usability~\cite{arxiv17:ray}. 

In this paper, we extend the TensorFlow runtime to enable automatic selection of intra-op parallelism for each operation and schedule operations for better performance of co-running operations. 
Our runtime is driven by performance models which aim to predict performance for operations with various intra-op parallelism. 
We explore two performance modeling methods. 
In the first method, we employ regression models that use performance events collected by hardware counters as input features. 
The runtime collects those features by dynamically profiling operations for a few times with different intra-op parallelisms. 
Our results show that this method does not provide good prediction accuracy and cannot effectively guide the selection of intra-op parallelism for operations. 
In the second method, we use a hill climbing algorithm to explore the shortest execution time and corresponding number of threads for each operation. 
Leveraging the execution history of the hill climbing approach, we predict the execution time of operations with various intra-op parallelisms. 
The hill climbing approach is very lightweight and causes high prediction accuracy (95\% with an appropriate configuration).

Based on the hill climbing algorithm-based performance model, our runtime system explores a set of scheduling strategies. In particular, we avoid frequent change of operation concurrency, because that causes performance loss due to cache thrashing and thread management overhead. We co-run multiple operations based on the performance model to maximize hardware utilization. We also leverage hyper-threading to allow multiple operations to share the same physical cores to improve system throughout. 

The major contributions of the paper is the following:

\begin{itemize}
\item We explore performance modeling to predict performance of operations with various intra-op parallelism; Our performance model is highly accurate and lightweight, hence can be effectively leveraged by the runtime to decide intra-op and inter-op parallelism of operations.

\item We study scheduling strategies that co-run operations to improve hardware utilization and system throughput.

\item We extend the TensorFlow runtime and demonstrate a significiant performance benefit of our approach. Comparing with the default configuration of intra-op and inter-op parallelisms for NN training in TensorFlow, our approach achieves 36\% performance improvement on average (up to 49\%) for 
four
NN models, and achieves performance close to the optimal one manually obtained by users. 
\end{itemize}

\section{Background}
\label{sec:background}

\subsection{Neural Network Training}

NN training could be expensive, because it is an iterative process involving large training data sets. 
Many NN take hours or even days for training, even on the state-of-the-art GPU~\cite{he2016deep}. 
Although using GPU to train neural network is common, using multi/many-core processors (e.g., Intel Knights Landing) to train neural network is also becoming common~\cite{
you2017scaling, kurth2017deep}. 


Training an NN often involves a large number of iterative steps (thousands and even millions of steps). In each step, a batch of samples is fed into the NN. 
Except the first step which is often used for performance profiling to determine appropriate data layout~\cite{sc16:li}, initialize data based on device configuration, and estimate performance by a cost model empirically or analytically~\cite{tensorflow2015-whitepaper},
all other steps have the same computation and memory access patterns. Performance of each step (particularly execution time and the number of main memory accesses) remains stable across steps.
The above characteristics allow us to build performance models based on dynamic profiling of the first few steps and use the profiling results to improve performance of the following steps.

Note that the word ``performance'' in this paper refers to the execution time, not modeling accuracy of NN.


\subsection{Dataflow-Based Machine Learning Framework}
The state-of-the-art ML frameworks, such as TensorFlow, Caffe2 and MxNet, 
decompose an ML model into fine-grained operations. 
Similar to task-based parallel programming models~\cite{vandierendonck2011parallel}
such operation-based ML frameworks greatly improve hardware utilization and system throughput~\cite{tensorflow2015-whitepaper}. 
Within a training step of NN training, there can be tens of different operations, and each operation can be invoked hundreds of times, each of which is an \textit{operation instance}. Different instances of an operation can have different input data sizes.

TensorFlow allows users to control operation concurrency. The operation concurrency includes inter-op parallelism and intra-op parallelism.
However, such control of operation concurrency has to be manually decided by the user. Furthermore, the intra-op parallelism is enforced uniformly on all operations, ignoring the scalability difference between operations.


\begin{figure}[tb!]
	\centering
	\includegraphics[width=0.85\linewidth]{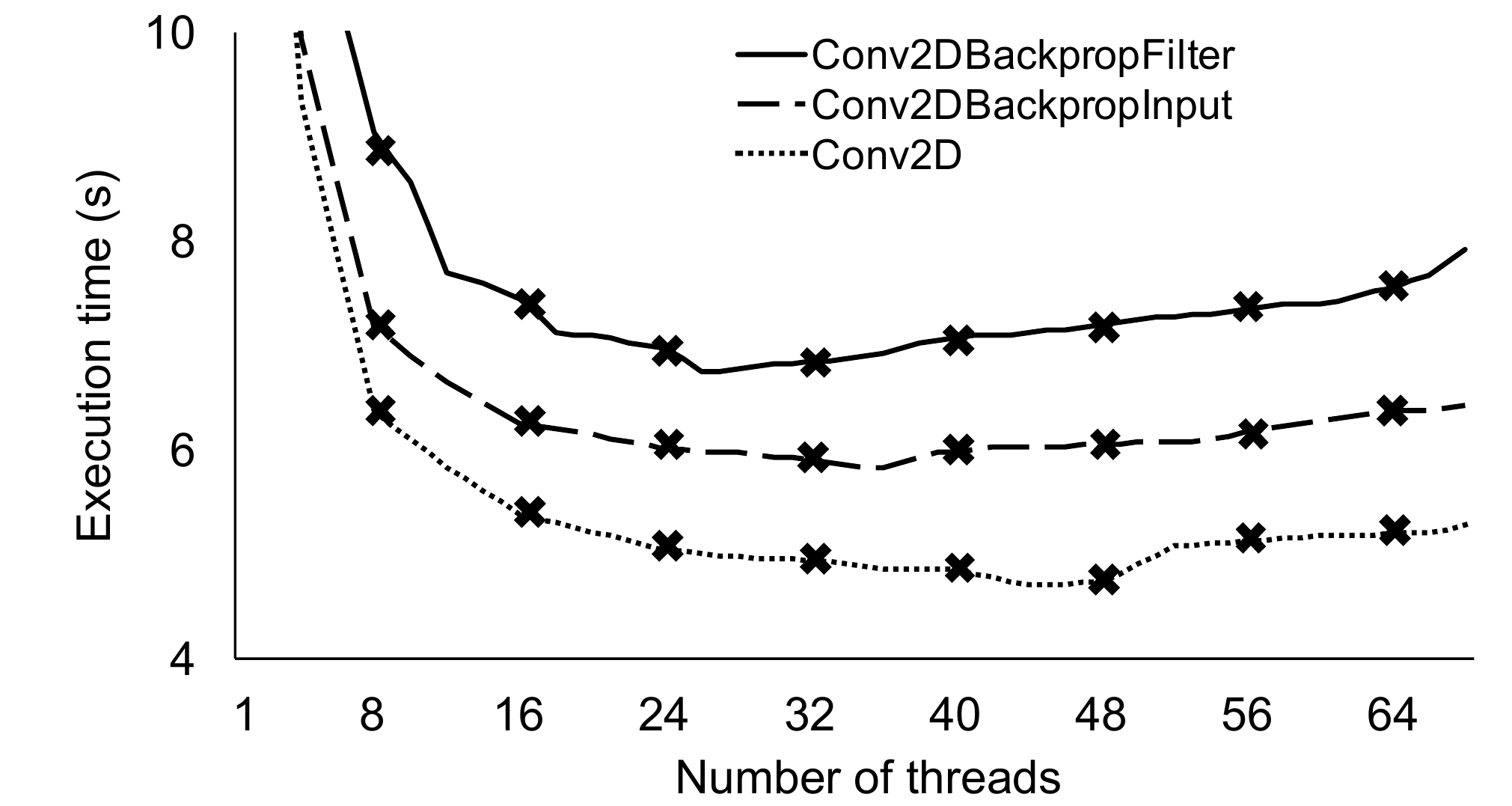}
	\caption{Performance variance of three operations with different intra-op parallelisms. The reported execution time is the total execution time of one thousand runs.
}   \vspace{-18pt}
    \centering
	\label{fig:threads_scale} 
    
\end{figure}

\begin{table}
\footnotesize 
\caption{Study the performance of NN models with different inter-op and intra-op parallelisms. The performance baseline for calculating speedup is the performance with the  configuration recommended by the TensorFlow programming guide (68 threads for intra-op parallelism and 1 for inter-op parallelism).} 

\label{tab:model_scale}
\centering

\resizebox{0.40\textwidth}{!}{%

\begin{tabular}{|c|c|c|c|c|c|}
\hline
\multicolumn{2}{|c|}{Parallelism} & \multicolumn{2}{c|}{ResNet-50} & \multicolumn{2}{c|}{DCGAN} \\ \hline
Inter-op        & Intra-op        & Time (ms)     & Speedup     & Time (ms)     & Speedup    \\ \hline
1               & 34              & 1414          & 0.98        & 484           & 1.21       \\ \hline
1               & 68              & 1382          & 1.00        & 524           & 1.00       \\ \hline
1               & 136             & 2257          & 0.61        & 1045          & 0.50       \\ \hline
2               & 34              & 1088          & 1.27        & 411           & 1.28       \\ \hline
2               & 68              & 1213          & 1.14        & 501           & 1.04       \\ \hline
2               & 136             & 4017          & 0.34        & 1238          & 0.42       \\ \hline
4               & 34              & 1169          & 1.18        & 434           & 1.21       \\ \hline
4               & 68              & 3048          & 0.45        & 565           & 0.93       \\ \hline
4               & 136             & 4782          & 0.29        & 1469          & 0.36       \\ \hline
\end{tabular}

}
\end{table}

\begin{table}
\footnotesize 
\caption{Study the impact of input data size on operation performance. The performance baseline for calculating performance variance is the performance with using 68 threads. The reported time is the total execution time of one thousand runs.}
\label{tab:op_profile}
\centering

\resizebox{0.49\textwidth}{!}{%

\begin{tabular}{|c|c|l|l|l|c|c|c|}
\hline
Operation Type                                  & \multicolumn{4}{c|}{Input data size} & Time (s) & Intra-Op Parallelism & Performance Variance \\ \hline
\multirow{3}{*}{Conv2DBackpropFilter} & \multicolumn{4}{c|}{(32,8,8,384)}    & 7.2     & 26             & 17.3\%                  \\ \cline{2-8} 
                                      & \multicolumn{4}{c|}{(32,17,17,384)}  & 11.1     & 42             & 10.2\%                  \\ \cline{2-8} 
                                      & \multicolumn{4}{c|}{(32,8,8,2048)}   & 20.3     & 68             & 0\%                     \\ \hline \hline
\multirow{3}{*}{Conv2DBackpropInput}  & \multicolumn{4}{c|}{(32,8,8,384)}    & 5.8     & 36             & 9.8\%                   \\ \cline{2-8} 
                                      & \multicolumn{4}{c|}{(32,17,17,384)}  & 8.7     & 56             & 2.3\%                   \\ \cline{2-8} 
                                      & \multicolumn{4}{c|}{(32,8,8,2048)}   & 19.6     & 68             & 0\%                     \\ \hline \hline
\multirow{3}{*}{Conv2D}               & \multicolumn{4}{c|}{(32,8,8,384)}    & 4.7     & 45             & 11.1\%                  \\ \cline{2-8} 
                                      & \multicolumn{4}{c|}{(32,17,17,384)}  & 7.4     & 63             & 3.5\%                   \\ \cline{2-8} 
                                      & \multicolumn{4}{c|}{(32,8,8,2048)}   & 14.8     & 66             & 2.0\%                   \\ \hline
\end{tabular}

}
\vspace{-10pt} 
\end{table}

\subsection{Motivation Examples}
\label{sec:motivation}
We study the performance characteristics of operations to motivate our concurrency control and operation scheduling. 
We perform our study from three perspectives: (1) Operation performance variance with different 
thread-level parallelisms; (2) Impact of the input data size on operation performance; (3) Performance impact of co-running operations.

\textbf{Hardware platform.}
We use Intel Knights Landing (KNL) processor (Xeon Phi 7250) as a manycore example in the rest of the paper. Several leadership supercomputers are based on KNL,
including Cori at Lawrence Berkeley National Lab and 
Theta at Argonne National Lab. 
KNL provides strong computation capabilities to train and deploy neural networks~\cite{tensorflow_nersc}. We use KNL processors at Cori for our study.

A KNL processor can contain 68 cores, each of which supports four hardware threads (in total 272 hardware threads).
68 cores are organized into 34 tiles (i.e., two cores per tile). Two cores in the same tile share a 1 MB L2 cache (the last level cache).
KNL has a 16GB on-package high-bandwidth memory. 
This memory can be configured as a transparent, direct-mapped hardware cache. This configuration is called ``cache mode''.
The cache mode is the most common mode in a KNL-based HPC. All the tests in this paper use the cache mode of KNL. With the cache mode, all data sets 
of NN models in our tests are placed in the high-bandwidth memory and there is no effect of non-uniform memory access (NUMA).

We use TensorFlow  (v1.9) in our study. 
We develop a performance profiling framework by leveraging TensorBoard~\cite{TensorBoard} and Intel VTune~\cite{reinders2005vtune} 
to collect timing and hardware counter information of operations.
The default intra-op and inter-op parallelisms in Tensorflow are set as the number of logical cores of the hardware platform (272 in KNL). However, the TensorFlow performance guide~\cite{performance_guide} recommends the user to set the inter-op parallelism as the number of sockets (which is one in our platform) and set the intra-op parallelism as the number of physical cores, which is 68 in our platform.



\subsubsection{Performance Variance with Different Concurrency}
We change the intra-op and inter-op parallelisms when running a couple of NN models (particularly ResNet-50 and DCGAN).  
Table~\ref{tab:model_scale} summarizes the results. There is a significiant performance variance across different cases. The default case (68 threads for intra-op parallelism and 1 for inter-op parallelism), which is our baseline,  does not result in the best performance. 
There is up to 28\% 
performance difference between the default case and the most performant case, as shown in Table~\ref{tab:model_scale}.


Furthermore, we change the number of threads to run individual operations (i.e., not the whole NN model). 
When running each operation with multiple threads, 
we put those threads with data sharing into the same tile for best performance (threads resident in the same tile share the last level cache). We do not use hyper-threading for the tests. When we run those individual operations, we develop a series of scripts to run them as standalone 
operations to avoid any performance interference between operations as in the NN model training.

Figure~\ref{fig:threads_scale} shows the execution times of three operations, $Conv2DBackpropFilter$, $Conv2DBackpropInput$ and $Conv2D$ with different number of threads. The three operations are common and among the most time-consuming operations in NN training~\cite{liu2018pim}. 
For those three operations, we use certain input data sizes in the NN model Inception-v3~\cite{szegedy2016rethinking}. 


Figure~\ref{fig:threads_scale} reveals that we achieve the best performance, when we use 26, 36 and 45 threads to run the three operations respectively. 
There is up to 17.3\% performance difference between the default concurrency (i.e., 68 threads) and the best case.
The scalability of the three operations with the given input data size is limited on KNL due to thread spawning overhead and non-parallelizable code regions. 

\textbf{Observation 1.} The intra-op parallelism must be chosen differently for different operations, in order to achieve the best performance of individual operations.

\subsubsection{Impact of Input Data Size}
An NN model can involve many instances of an operation in a training step. Different instances of the operation can use different input data sizes. 
For example, in Inception-v3,  the operation $Conv2DBackpropFilter$ has 42 instances in a training step, each of which uses different input data sizes. 


We study three operations from Inception-v3,  which are $Conv2DBackpropFilter$, $Conv2DBackpropInput$ and $Conv2D$. We change the input data sizes of the three operations. For each input data size, we change the intra-op parallelism to find the best performance. 
Table~\ref{tab:op_profile} shows the results. 
The table shows that as we change the input data sizes, we need to use different numbers of threads to achieve the best performance. For example, for $Conv2DBackpropFilter$ with the input data size   $par\_input$ as (32,8,8,384), we need to use 26 threads to achieve the best performance, while with the input data size $par\_input$ as (32,17,17,384) and $par\_input$ (32,8,8,2048), we need to use 42 and 68 threads, respectively.



\textbf{Observation 2.} The best concurrency (in terms of the optimal number of threads per operation)
changes, as we change the input data size of the operation.

\subsubsection{Co-Running Operations}
We study the performance of co-running operations. 
We use three strategies to run two operations. First, we run them in serial, and each operation uses 68 threads.
This strategy would be used by the TensorFlow runtime by default. Second, we leverage two hardware threads (hyper-threading) in each core to allow the two operations to co-occupy 68 cores (i.e., each operation uses 68 cores and there is one hardware thread per core for each operation). 
Third, we evenly partition 68 cores between the two operations (i.e., each operation uses 34 core; only one hardware thread per core). The performance of co-running the two operations is the time span from launching them to finishing both of them.

Table~\ref{tab:corun} summarizes the results of co-running $Conv2DBackpropFilter$ and $Conv2DBackpropInput$.
The input size for the operations is $par\_input$ (32,8,8,2048).
Given this input size, the number of threads to achieve the best performance for the two operations is 68.

\begin{table}
\footnotesize 
\caption{Co-running two operations with three strategies. The performance baseline for calculating speedup is performance of serial execution of two operations. The reported time is the total execution time of one thousand runs.
} 

\label{tab:corun}
\centering

\resizebox{0.45\textwidth}{!}{%

\begin{tabular}{|c|c|c|c|}
\hline
Strategies                  & \#Threads & Time (s) & Speedup \\ \hline
Serial execution        & 68        & 41.1     & 1       \\ \hline
Co-run with hyper-threading & 68+68     & 39.9     & 1.03    \\ \hline
Co-run with threads control & 34+34     & 29.8     & 1.38    \\ \hline
\end{tabular}

}
\vspace{-18pt}
\end{table}

Table~\ref{tab:corun} reveals that the third strategy achieves the best performance. Comparing with the first strategy, the third one has 38\% performance improvement, although individual operations have performance loss (25\% and 17\% for $Conv2DBackpropFilter$ 
and 
$Conv2DBackpropInput$
respectively). Hyper-threading (the second strategy) is helpful in this study: we have 3\% performance improvement (comparing with the first strategy).



\textbf{Observation 3.} Co-running operations are helpful for overall performance improvement, even though individual operations may have performance loss when co-running them.

\section{Design}
\vspace{-5pt}
\subsection{Overview}
\begin{figure}[tb!]
	\centering
	\includegraphics[width=0.99\linewidth]{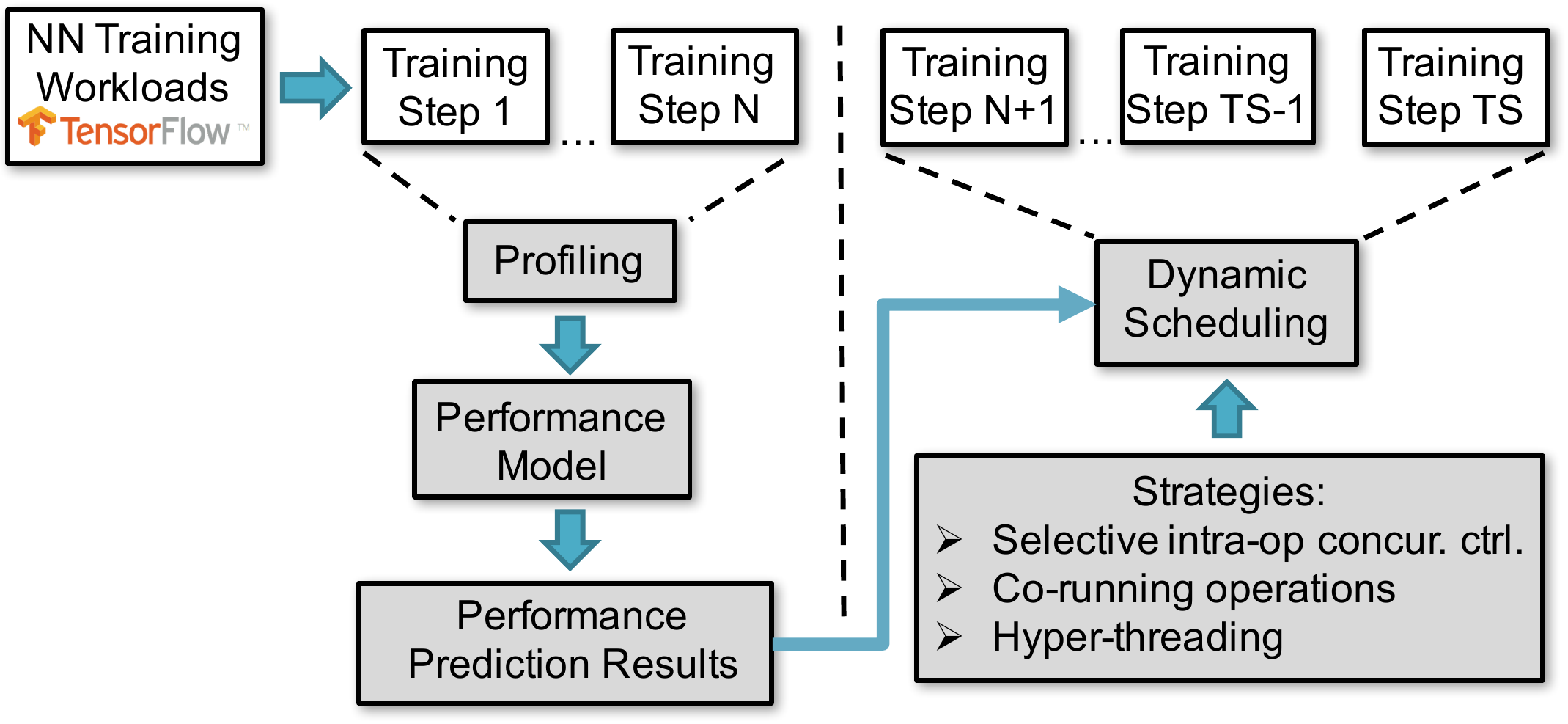}
    \vspace{-5pt}
	\caption{Our runtime framework and its workflow. The notation ``TS'' is the total number of training steps.} 
    \centering
	\label{fig:general_framework} 
    \vspace{-18pt}
\end{figure}

\vspace{-5pt}
\label{sec:overview}
Our motivation examples demonstrate the necessity of dynamically changing concurrency (intra-op and inter-op parallelisms) and scheduling operations to reduce training time. Driven by the motivation examples, we use a performance model-driven approach 
to extend the TensorFlow runtime. Figure~\ref{fig:general_framework} generally depicts our runtime and its workflow.

In particular, we explore two performance models to predict performance (execution time) of each operation with different intra-op parallelisms. 
We study the performance modeling accuracy and model portability across architectures and operation implementations.
We further extend the TensorFlow runtime to schedule operations to enable co-running operations.

Our two performance models are based on dynamic profiling. The performance models use a few training steps (the profiling steps) to profile operations and then make performance prediction on operations with various intra-op parallelisms.

Our first performance model uses machine learning models (a set of regression models) to make the performance prediction. Those models use characteristics of operations as input features.
We characterize computation and memory access patterns of operations by collecting performance events during the profiling steps. However, the machine learning models have low prediction accuracy because execution times of some operations are short and collecting performance events with hardware counters within such short times is not accurate.
Furthermore, the regression models are architecture dependent and have to be re-trained on a new platform. 

Our second performance model is based on the hill climbing algorithm~\cite{hill_climbing}.
Our hill climbing algorithm aims to find the best performance (the shortest execution) and corresponding number of threads to run an operation with a given input data size.
The algorithm starts with a certain number of threads to run the operation, then attempts to find another number of threads with a shorter execution time by making an incremental change to the number of threads. If the change produces a shorter execution time, another incremental change is made to the number of threads until no further execution time is reduced. 
When running the hill climbing algorithm, the runtime tests a few cases (i.e., the profiling cases) and measure their execution times. To predict the performance of any untested case, we use linear interpolation to predict the performance of the untested case based on the measured performance of two profiling cases.
The hill climbing algorithm-based performance model is lightweight and accurate. Different from the first performance model, it is architecture-independent and require no knowledge of operations. Hence, we use the second performance model to guide the runtime.  



Our runtime system decides (1) the optimal intra-op parallelism for each operation and (2) which operations to co-run to increase system utilization.
The performance model determines the optimal number of threads for an operation that results in the shortest execution time.
To avoid frequent changes of operation concurrency, which may lead to sub-optimal overall performance, we optimize operation concurrency for the largest input size, which yield overall better performance.
Furthermore, our runtime decides which operations should co-run and how they should interleave.
We analyze a set of candidate execution scenarios and select the one that best suits the current execution flow to increase overall system throughput and hardware utilization under the constraints of available computing resources.
Our runtime system also leverages hyper-threading to allow multiple operations to share the same physical cores to improve system throughout.
In the following, we describe our design in detail.

\vspace{-5pt}
\subsection{Regression Model-Based Performance Model}
\label{sec:perf_model}

Our first performance models, which are regression-based, predict performance 
for 68 cases, each of which has a specific number of threads.
For each case, we have a performance model to make prediction (68 performance models in total).
For each prediction case, there is only one thread per core. 
We do not predict the cases with multiple threads per core (i.e., using hyper-threading), because hyper-threading often causes performance slowdown, when running a single operation.
 
Among the 68 prediction cases, 34 of them have at most one thread in each tile. In other words, those 34 cases do not have any cache sharing between threads. The remaining 34 cases have either two threads or no threads in each tile. 
In other words, those 34 cases have cache sharing between threads. 
For those 34 cases, we only use even number of threads. We do not consider odd number of threads, because that makes some tile have only one thread, causing load imbalance between tiles. 

Note that we use 68 performance models to predict 68 cases. We do not try to build a single model to predict performance of the optimal case (the case with the shortest execution time), because the runtime system needs to know the performance of many cases to decide which operations to co-run.
We also do not try to build a single model to predict performance of 68 cases, because of the complexity of model training and low prediction accuracy (as low as 25\% according to our study). 

Each performance model collects a set of workload features as model input, using a few training steps. 
We consider \textit{thread affinity} while collecting features.
Thread affinity decides the binding between threads and cores.
For those threads with large data sharing, we want to bind them into the same tile, such that those threads can reuse data in the L2 cache of the tile. Given the number of threads per tile and total number of threads to run an operation, different thread affinity can result in different performance. Our model aims to predict performance with the best thread affinity.

When running operations with a specific number of threads and measuring the execution times of those operations, we carefully choose which two threads should share a tile.
In particular, we put the threads with continuous IDs into the same tile. For example, threads 1 and 2 share a tile, and threads 3 and 4 share a tile. 
This method is based on the following observation: The multi-threading mechanism (i.e., OpenMP) used in TensorFlow on KNL
is implemented in the Intel MKL-DNN library, and 
this mechanism parallelizes operations by assigning iterations of the major computation loop to threads in order, and neighbor iterations tend to access the same data set, hence the threads with continuous IDs that work on the neighbor iterations tend to have data sharing. 

The above method provides a lightweight and practical solution to enforce thread affinity for best performance. There are other solutions that involve compiler and runtime analysis~\cite{mazouz2011performance}, but they are expensive.



\noindent\textbf{Feature Selection.} 
We use performance events collectible by hardware counters, 
and the execution time of the operation, as features. In total there are 27 features.

On KNL, there are 26
performance events collectible by hardware counters. Using all of them as features is problematic due to the following reasons. First, those performance events cannot be collected at the same time. We need at least four training steps to collect those events separately,  which increases the number of training steps for profiling. Second, some features are not informative, discriminating and independent. For example, the number of branch instructions and number of conditional branch instructions 
are correlated and redundant, and should not be selected together.




We employ the decision tree estimator to select features. 
We choose four features: number of CPU cycles, number of last level cache misses, number of last level cache accesses and number of level 1 cache hits.
We also normalize the numbers of performance events by the total number of instructions to make the feature values independent of total number of instructions.
The normalization makes the performance model usable for workloads with different number of instructions.


\noindent\textbf{Feature collection.}
We collect features using $N$ sample cases. Each sample case uses a specific number of threads to run a training step. All operations in this training step use the same number of intra-op parallelism. In the training step, we run the operations in serial to avoid performance interference among multiple operations and ensure accuracy of feature collection.   

We choose sample cases by evenly sampling the search space of possible intra-op parallelisms with the consideration of cache sharing. Using those sample cases is meant to be representative of all cases. 

To decide the number of sample cases ($N$), we change $N$ to study its impact on modeling accuracy. The results are summarized in Table~\ref{tab:performance_model}. The results reveal that $N$ has a significant impact on modeling accuracy, but a large $N$ is not helpful for improving modeling accuracy. Also, using a large $N$ can cause large runtime overhead, because of frequent counting performance events for a large number of operations.
In our test with ResNet-50, when $N=16$, the runtime overhead is up to 20\%. 

\noindent\textbf{Regression models.}
We experiment with ten regression models and compare their accuracy,  including random forest, k-nearest neighbors, gradient boosting, $\epsilon$-support vector machine regressions ($\epsilon$-SVR) with linear, poly and RBF kernels, decision tree, Bayesian automatic relevance determination (ARD), ordinary least squares (OLS), passive aggressive regression (PAR), multiple layer perceptron (MLP) with sgd, lbfgs and adam kernels and Theil Sen Regression (TSR).


\begin{table}
\footnotesize 
\caption{
Prediction accuracy of a set of regression models.
} 

\label{tab:performance_model}
\centering

\resizebox{0.49\textwidth}{!}{%

\begin{tabular}{|c|c|c|c|c|c|c|}
\hline
\#Sample ($N$)            & Metrics  & Gradient Boosting & K-Neighbors & TSR   & OLS   & PAR   \\ \hline
\multirow{2}{*}{1}  & Accuracy & 61\%              & 56\%        & 37\%  & 27\%  & 18\%  \\ \cline{2-7} 
                    & $R^2$    & 0.961             & 0.818       & 0.779 & 0.981 & 0.196 \\ \hline
\multirow{2}{*}{4}  & Accuracy & 57\%              & 67\%        & 17\%  & 21\%  & 14\%  \\ \cline{2-7} 
                    & $R^2$    & 0.957             & 0.592       & 0.539 & 0.951 & 0.175 \\ \hline
\multirow{2}{*}{8}  & Accuracy & 51\%              & 56\%        & 26\%  & 31\%  & 18\%  \\ \cline{2-7} 
                    & $R^2$    & 0.972             & 0.589       & 0.965 & 0.977 & 0.177 \\ \hline
\multirow{2}{*}{16} & Accuracy & 34\%              & 26\%        & 13\%  & 14\%  & 11\%  \\ \cline{2-7} 
                    & $R^2$    & 0.959             & 0.585       & 0.852 & 0.892 & 0.159 \\ \hline
\end{tabular}

}
\vspace{-18pt}
\end{table}

\noindent\textbf{Training Data Set.}
For training data set, we collect operation information from three common NN models with TensorFlow (particularly ResNet-50
with CIFAR-10 dataset, 
 DCGAN~\cite{DBLP:journals/corr/RadfordMC15} with MNIST dataset and    
Inception-v3 with ImageNet dataset.
To increase training data set, we vary batch size from 16 to 256. 
When we run those operations in the three NN models
, we develop scripts to run them as standalone operations, similar to what we do in the motivation examples (Section~\ref{sec:background}).

\noindent\textbf{Model Testing.}
We test model accuracy with DCGAN.
Table~\ref{tab:performance_model} shows the results. We use two metrics, modeling accuracy and $R^2$ (the coefficient of determination). 
The modeling accuracy is defined as $1 - \frac{1}{n_{t}}\sum\left | \frac{\hat{y}_{i} - y_{i} }{y_{i}}\right|$
where $n_{t}$ is the size of the test data set, and $\hat{y}_{i}$ and $y_{i}$ are the predicted and actual execution times, respectively.

Table~\ref{tab:performance_model} shows that the regression-based performance models do not present good accuracy for the selection of operation concurrency. Using the most accurate regression model (k-neighbors) to direct 
NN model training (ResNet-50 in particular), we have 
performance loss (30\%).

We attribute those prediction inaccuracy to possible inaccuracy in hardware counters to collect performance events. 
Using hardware counters 
can be inaccurate.
Furthermore, the regression model-based performance models are architecture-dependent. The regression models need to be re-trained on a platform with different hardware counters. 


Because of the above reasons, we propose to use a hill climbing algorithm to direct the selection of intra-op parallelism for operations.

\vspace{-3pt}


\subsection{Hill Climbing Algorithm-Based Performance Model}
\label{sec:hill_climbing}

We describe our hill climbing algorithm as follows.
Similar to the regression-based performance models, we use $N$ training steps to run operations in serial with different number of threads. 
In particular, we first use one thread to run each operation and measure execution time in one step. Then, we increase the number of threads by $x$ (named as \textit{interval}) to run each operation and measure its execution time. By increasing the number of threads, the execution time can decrease. We further to increase the number of threads by $x$ in the following steps, until one of the following two cases happens: (1) the execution time increases; (2) we reach the maximum number of cores to run threads.
If (1) happens, then we stop changing the number of threads for this operation and claim that we find the best number of threads to run the operation in the last time step. 
If (2) happens, then the best number of threads to run the operation is the maximum number of cores.

We consider thread affinity in the above hill climbing algorithm. 
In particular, given a specific number of threads to run an operation, we run the operation twice with two training steps: one step with cache sharing between threads, and the other without cache sharing between threads. 

The output of the above hill climbing algorithm includes not only the shortest execution time and corresponding number of threads, but also the execution time of those sampling cases in the $N$ training steps. 
To predict the performance of those cases that are not tested in the $N$ steps, we simply use linear interpolation. For example, if we measure the execution time of using one and four threads for an operation ($x=3$ in this example), then the execution time of using two and three threads will be approximated based on a linear interpolation between the execution times of using one and four threads.

$N$ (the number of training steps to run sample cases) is related to $x$.
Assuming that the maximum number of cores is $C$, then $N$ is at most $C/x \times 2$ (we have ``2'', because we consider both cache-sharing and no-cache-sharing cases.) 

\begin{table}
\footnotesize 
\caption{Performance prediction accuracy for four NN models based on the hill climbing-based performance model.}

\label{tab:hill_climb_prediction}
\centering

\resizebox{0.4\textwidth}{!}{%

\begin{tabular}{|c|c|c|c|c|}
\hline
             & \multicolumn{4}{c|}{Intervals}        \\ \hline
Models       & 2       & 4       & 8       & 16      \\ \hline
ResNet-50    & 98.13\% & 95.45\% & 83.42\% & 31.12\% \\ \hline
DCGAN        & 97.16\% & 94.43\% & 51.54\% & 10.14\% \\ \hline
Inception-v3 & 97.91\% & 94.22\% & 73.21\% & 21.21\% \\ \hline
LSTM         & 95.56\% & 90.45\% & 41.34\% & 11.03\% \\ \hline
\end{tabular}

}
\vspace{-18pt}
\end{table}

\textbf{Performance prediction accuracy.}
We run ResNet-50, DCGAN, Inception-v3 and LSTM.
 and use the hill climbing algorithm-based performance model to predict performance of those cases not executed in the $N$ steps.
We change $x$ from 2, 4, 8 to 16.  
Table~\ref{tab:hill_climb_prediction} shows the prediction accuracy. 
The prediction accuracy is the average prediction accuracy for all operations. 
In general, we achieve very high prediction accuracy (up to 98.13\% with $x=2$ and 95.45\% with $x=4$), much higher than regression model-based performance models (Section~\ref{sec:perf_model}).


\textbf{Discussion.}
Using the hill climbing algorithm has two potential problems.
First, the ``shortest execution time'' found by the hill climbing algorithm may be a ``local optimum'' solution, not a ``global optimum'' solution. However, after extensive evaluation of operation performance (1025 operations in four NN models) with different number of threads, we observe that the local optimum is always the global optimum. As the number of threads changes, the variance of execution time is shown as a convex function. 

Second, if the interval $x$ is large, it is possible that the hill climbing algorithm may skip the optimum. For example, assuming that the hill climbing algorithm has tested the case of eight threads, $x=4$, and the optimum is the case of 10 threads, then the hill climbing algorithm will only test the case of 12 threads and skip the optimum. The case of 12 threads is incorrectly selected as the optimum. However, our evaluation reveals that the optimum found by the hill climbing is pretty close to the real optimum. 
With the evaluation of four NN models (ResNet-50, DCGAN, Inception-v3 and LSTM) and $x=4$, the performance difference between the two optimums is less than 2\%.

In conclusion, the performance model based on hill climbing is a practical and effective approach for performance profiling and prediction. Comparing with the regression model-based performance models, the hill climbing has the following advantages: (1) No need of performance model training; 
(2) architecture independence; (3) no need of considering operation characteristics, hence can accommodate any future change of operations in TensorFlow; and (4) better accuracy.

\vspace{-2pt}

\vspace{-5pt}
\subsection{Runtime Scheduling}
\label{sec:runtime}

The runtime decides (1) intra-op parallelism for each operation, and (2) which operations to co-run. The existing runtime system in TensorFlow employs a first-in-first-out policy to schedule operations: The operations that are ready to run are simply executed in the order they put into the operation queue. All operations use the same intra-op parallelism and inter-op parallelism defined by the user before the training starts. Such scheduling strategy loses performance without sufficient consideration of operation scalability and hardware utilization. Our runtime avoids this problem and schedules operations based on the following strategies.


\textbf{Strategy 1: Deciding intra-op parallelism for individual operations based on the performance model.}
After running the hill climbing algorithm in the first few steps, the runtime runs each instance of each operation using the number of threads that can lead to the shortest execution time. 
This indicates that different operations may use different number of threads;
This also indicates that different instances of an operation with different input data sizes may also use different numbers of threads.

\textbf{Strategy 2: Avoiding frequent change of operation concurrency.}
In practice, Strategy 1 
might not lead to better performance than the execution with the default TensorFlow configuration. 
The reason is because of frequent change of operation concurrency, which causes cache thrashing and large thread management overhead (e.g., thread spawning or binding to cores). In Strategy 2, the runtime avoids frequent change of operation concurrency. In particular, the operation, no matter what input data size it uses, always use the same number of threads, but different operations can still use different number of threads. 
The number of threads to run the operation is determined by
the operation instance with the largest input data size (the most time-consuming instance), such that the execution time of this operation instance is the shortest.

\textbf{Strategy 3: Co-running operations to maximize hardware utilization.}
To decide which operations should co-run and how they should co-run, we use the following algorithm. 
 For any operation ready to run, we use three different numbers of threads as \textit{candidates} to run the operation (The ``three'' is an empirical number). 
The three candidates 
should be the most performant ones (i.e., the ones with the shortest execution times). 
Whenever some physical cores are idling, either because an operation is just finished or because we just start the training, we examine  those operations ready to run. For each of those operations, we check if any of its three candidates can fit into the idling cores without decreasing system throughput. 
We decide whether system throughput will be decreased by ensuring that the candidate does not take longer execution time than ongoing operations in busy cores.


For example, an operation ready to run has three candidates, which are (1) using 18 threads (no cache sharing) that takes 1.5 seconds; (2) using 20 threads (no cache sharing) that takes 1.3 seconds; and (3) using 16 threads (no cache sharing) that takes 2.1 seconds. We have 20 idling cores, and the remaining 48 cores run an ongoing operation that needs 1.9 seconds to finish. We choose the candidate (1) to co-run with the ongoing operation, because it takes shorter execution time than the ongoing operation (1.5 vs. 1.9 seconds), and can fit into the 20 idling cores.
We do not use 20 threads to fit into the 20 idling cores, because using 18 threads can release two idling cores to run another operation and we want to \textit{maximize operations co-running to increase system throughput}. 
An argument to support using 20 threads is that we finish the operation earlier and then run another operation. However, 
according to our experience, maximizing operations co-running (using 18 threads) is helpful to system throughput and hence more beneficial for performance. Note that the above execution times for operations are predicted based on the performance model.

If we cannot find any operation that can fit into the idling cores without decreasing system throughput, we choose the most time-consuming operation to run.

Strategy 3 should not conflict with Strategy 2. If the number of threads to run an operation based on Strategy 3 is quite different from the number of threads chosen by Strategy 2 
(the difference in the number of thread is larger than 2 and ``2'' is an empirical value),
 then we will use the number of threads chosen by Strategy 2 to run the operation. 
This method avoids disruptive changes to intra-op parallelism for each operation.

Strategy 3 is lightweight and can make a quick decision on how to co-run operations, such that the runtime overhead is small. Based on our profiling on four neural networks (ResNet-50, DCGAN, Inception-v3 and LSTM), we seldom have more than five operations ready to run. Hence, Strategy 3 does not need to explore a lot of operations to make the decision.

\textbf{Strategy 4: Leveraging hyper-threading to run multiple operations.}
Some scalable operations can take all 68 cores and never allow any operation to co-run. However, we find that running small operations using hyper-threading along with the time-consuming, scalable operations can be beneficial for performance. This means that the small operations share physical cores with the time-consuming operations, enabling another type of co-run.

At runtime, when the runtime finds an operation using 68 cores, the runtime then tries to co-run small operations. The small operations are defined as those operations that have shortest serial-execution time in the operation-ready queue.



\textbf{Putting all together.}
The runtime uses Strategies 1-2 to decide the number of threads to run for each operation based on the performance model. This can be done right after running the hill climbing algorithm in the first few training steps (the profiling steps). After that, the runtime decides how to co-run operations based on Strategies 3-4.  
The runtime repeatedly uses the four strategies until all operations are finished.
Note that to minimize runtime overhead, some decisions based on Strategy 3 to co-run operations can be reused without repeatedly running Strategy 3.

\textbf{Discussion.}
Our performance model is used to predict performance for individual operations, and does not capture performance interference between operations when co-running them. 
Hence, when we use the performance model to direct operations to co-run, the performance loss of individual operations can be unexpected low because of performance interference. Our runtime can record such cases and avoid co-running such operations in the future train steps. In practice, we do not find significant performance slowdown in individual operations when co-running them.

\vspace{-3pt}

\section{Evaluation}
\label{sec:evaluation}
\begin{figure*}[tb!]
  \centering
  \includegraphics[width=0.99\linewidth]{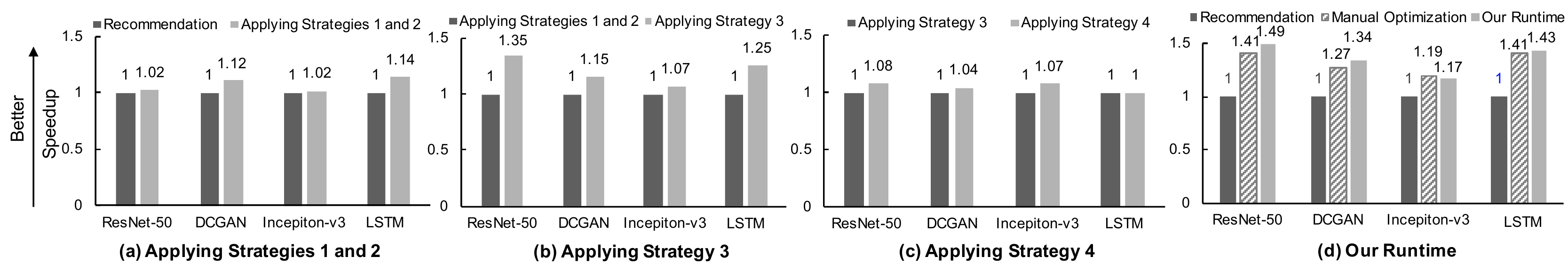}
%
  \caption{Quantifying the contribution of the four strategies. Comparing the performance of our runtime, manual optimization, and the recommendation by TensorFlow.} 
  \centering
  \label{fig:performance}
\end{figure*} 

\begin{table*}
\footnotesize 
\caption{Performance improvement of the top five most time-consuming operations in four NN models by recommendation and by applying Strategies 1 and 2. The performance baseline for calculating speedup is the performance with the configuration recommended by the TensorFlow programming guide (68 threads for intra-op parallelism and 1 for inter-op parallelism).}
\label{tab:strategies_1_2}
\centering

\resizebox{1.0\textwidth}{!}{

\begin{tabular}{|c|c|c|c|c|c|c|c|}
\hline
Operations           & \multicolumn{2}{c|}{Execution Time (ms)}     & Speedup & Operations           & \multicolumn{2}{c|}{Execution Time (ms)}     & Speedup \\ \hline
                     & Recommendation & Applying Strategies 1 and 2 &         &                      & Recommendation & Applying Strategies 1 and 2 &         \\ \hline
\multicolumn{4}{|c|}{ResNet-50}                                               & \multicolumn{4}{c|}{DCGAN}                                                    \\ \hline
Conv2DBackpropFilter & 158            & 146                         & 1.08    & Conv2DBackpropInput  & 164            & 144                         & 1.14    \\ \hline
InputConversion      & 131            & 122                         & 1.07    & Conv2DBackpropFilter & 133            & 110                         & 1.21    \\ \hline
Tile                 & 107            & 105                         & 1.02    & ApplyAdam            & 84             & 72                          & 1.17    \\ \hline
Mul                  & 103            & 100                         & 1.03    & BiasAddGrad          & 26             & 23                          & 1.17    \\ \hline
ToTf                 & 79             & 78                          & 1.01    & FusedBatchNorm       & 15             & 14                          & 1.03    \\ \hline
\multicolumn{4}{|c|}{Inception-v3}                                            & \multicolumn{4}{c|}{{LSTM}}                                                     \\ \hline
AvgPool              & 759            & 730                         & 1.04    & SparseSoftmaxCross   & 11.71          & 8.76                        & 1.34    \\ \hline
Tile                 & 539            & 532                         & 1.01    & BiasAddGrad          & 2.03           & 1.98                        & 1.03    \\ \hline
Conv2DBackpropFilter & 479            & 475                         & 1.01    & Mul                  & 1.36           & 1.09                        & 1.25    \\ \hline
MaxPooling           & 455            & 422                         & 1.08    & AddN                 & 1.02           & 0.87                        & 1.17    \\ \hline
InputConversion      & 416            & 413                         & 1.01    & MatMul               & 0.95           & 0.93                        & 1.02    \\ \hline
\end{tabular}

}
\vspace{-12pt} 
\end{table*}

\subsection{Experiment Setup}
\textbf{Training models, data set and framework.} We employ CIFAR-10, MNIST, ImageNet and PTB training dataset for ResNet-50, DCGAN, Inception-v3 and LSTM respectively. The batch sizes of ResNet-50, DCGAN, Inception-v3 and LSTM are 64, 64, 16 and 20, respectively. 
We adopt TensorFlow (v1.9) as our NN training framework. 
We use the implementation of ResNet-50, Inception-v3 and LSTM from the TensorFlow software package~\cite{nn_models_github} and DCGAN from~\cite{dcgan_github}. In TensorFlow, the default intra-op and inter-op parallelisms are set as the number of logical cores of the hardware platform (272 in KNL). As discussed in Section~\ref{sec:background}, the TensorFlow performance guide recommends to set the inter-op parallelism as the number of sockets (which is one in our platform) and set the intra-op parallelism as the number of physical cores, which is 68 in our platform. Since the performance of the TensorFlow default configuration is much worst (more than 10 times slower) than the recommended configuration from the TensorFlow performance guide, we use the \textit{recommended configuration} as the baseline in our evaluation. The performance with the recommended configuration is annotated as ``Recommendation'' in Figure~\ref{fig:performance} and Table~\ref{tab:strategies_1_2}.

The performance reported in this section is the execution time of one training step. Recall that the performance of one training step remains stable across training steps, hence the execution time of one training step is good for performance evaluation. 
In addition, there is no accuracy loss in NN models with our runtime, because our runtime does not make any change to the input data sizes of operations, does not change any NN model parameters, and does not violate any dependency between operations.


\textbf{Hardware platform.} 
We use a machine with an Intel Knights Landing (KNL) processor (Xeon Phi 7250) at the Cori supercomputer at Lawrence Berkeley National Lab as our test platform. 
Section~\ref{sec:motivation} has more details for KNL.

\textbf{Controlling intra-op parallelism.}
On Intel KNL, TensorFlow uses operations implemented 
in both MKL-DNN and Eigen. 
Dynamically changing intra-op parallelism for those operations implemented in the Eigen causes large runtime overhead (larger than 10\%), because the Eigen decomposes an operation into a large number of tasks, and changing intra-op parallelism of an operation causes frequent task-pushing into and task-popping out of a queue associated with each thread. MKL-DNN uses OpenMP threads to parallelize operations, and there is negligible overhead to change intra-op parallelism for those operations implemented in MKL-DNN.
Hence, in our evaluation, we only change intra-op parallelism for those operations implemented in MKL-DNN. Those operations take more than 70\% of total NN training time.

To enable dynamic change of intra-op parallelism for a few operations (e.g., batch\_normalization in DCGAN), we have to make small changes to the operation implementation.
For example, we have to allocate a larger memory space for some variables during operation initialization.
However, the changes are minor and have no impact on operation performance. 

In general, the implementation of our runtime incurs limited overhead (less than 1\%). Also, the number of profiling steps is small (less than 0.05\% of total training steps). Hence, the profiling overhead is negligible.

\vspace{-3pt}

\subsection{Results}
\label{sec:results}
\vspace{-3pt}






Figure~\ref{fig:performance}.d compares the performance of our runtime system with that of the recommended TensorFlow configuration (labeled as ``recommendation'') and of manual optimization. 
For manual optimization, we manually change intra-op and inter-op parallelisms uniformly enforced on all operations, aiming to find the best configuration. The manual optimization is not a scalable solution, because we have to exhaustively test every possible combination of intra-op and inter-op parallelisms to find the best configuration.

Figure~\ref{fig:performance}.d reveals that our runtime leads to the best performance in all tests. Our runtime performs at least 17\% (Inception-v3) and up to 49\% (ResNet-50) better than the recommendation. Our runtime performs even better (at least 2\%) than the manual optimization for three NN models (ResNet-50, DCGAN and LSTM), and performs similar (2\% worse) to the manual optimization (Inception-v3).

The above results demonstrate the superior performance of our runtime system. To further understand the performance contributions of four runtime strategies, we apply them one by one. The results are shown in Figure~\ref{fig:performance}.a-Figure~\ref{fig:performance}.c.


\textbf{Applying Strategies 1 and 2 (concurrency control for individual operations)}
Figure~\ref{fig:performance} shows that applying Strategies 1 and 2 alone, we have 14\% performance improvement for LSTM, 12\% for DCGAN and 2\% for ResNet-50 and Inception-v3. 

Table~\ref{tab:strategies_1_2} shows the execution times of the top five most time-consuming operations of four NN models with the recommended TensorFlow configuration and with Strategies 1 and 2 in place. The table reveals that we have better performance for all operations, up to 34\% performance improvement.   

Some operations do not have performance improvement after applying Strategies 1 and 2, however these operations (e.g., Conv2D in ResNet-50) with our runtime can use less number of threads than with the recommendation, while achieving the same performance.
Using less number of threads introduces opportunities to co-run operations.


\textbf{Applying Strategy 3 (co-running operations).}
To isolate the effects of co-running operations from Strategy 4, we apply Strategy 3 after using Strategies 1 and 2 without Strategy 4.

Figure~\ref{fig:performance}.b shows the results. 
The performance reported in the figure is normalized by the performance of applying Strategies 1 and 2. 
By using Strategy 3, ResNet-50 achieves 35\% performance improvement. LSTM achieves 25\% performance improvement. DCGAN and Inception-v3 achieve 15\% and 7\% performance improvement, respectively. 

\textbf{Applying Strategy 4 (hyper-threading).}
To isolate the effects of Strategy 4 from other strategies, we apply Strategy 4 after applying Strategy 3 (this implicitly indicates that we also apply Strategies 1 and 2). 

Figure~\ref{fig:performance}.c shows the results. 
The performance reported in the figure is normalized by the performance of applying Strategies 3. ResNet-50, DCGAN, and Inception-v3 achieve 8\%, 4\%, and 7\% performance improvement, respectively. 
LSTM has no performance improvement, because almost no operation in LSTM needs all of cores to achieve best performance, hence there is few opportunity to apply Strategy 4.

\begin{figure*}[tb!]
  \centering
  \includegraphics[width=0.8\linewidth]{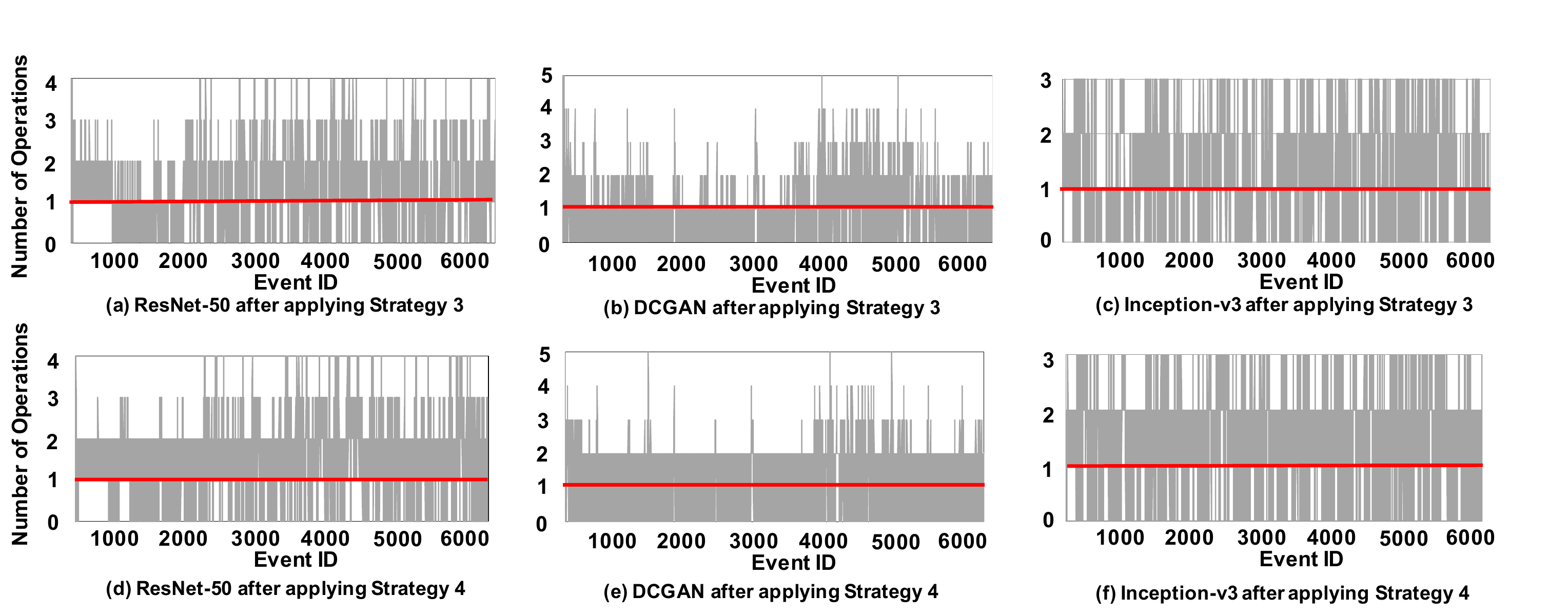}
  \vspace{-5pt}
  \caption{The variance of the number of co-running operations along with the NN model training. The figures (a), (b) and (c) do not have Strategy 4 (but have Strategy 3); The figures (d), (e) and (f) have Strategy 4 in place. The red 
lines in the figures are the inter-op parallelism recommended by TensorFlow.}
  \centering
  \label{fig:timeline}
\vspace{-18pt}  
\end{figure*}

To further study the effectiveness of Strategy 4, we record the number of co-running operations along with the NN training.
In particular, whenever there is an operation finished or launched, we record the number of co-running operations at the moment.
Finishing or launching an operation is an \textit{event}.
Figure~\ref{fig:timeline} shows the number of co-running operations whenever an event happens. There are a large number of events (sometimes millions of events), in just one training step. Presenting the number of co-running operations for all events makes the figure very intensive and difficult to read. Hence, we present the number of co-running operations for 6000 events in Figure~\ref{fig:timeline}.
The events happen in the middle of one step.
Figure~\ref{fig:timeline} does not show the results for LSTM, because there is no change in co-running operations after applying Strategy 4.

Figure~\ref{fig:timeline} shows that with Strategy 4 in place, the number of co-running operations is larger than that without Strategy 4 (but with Strategy 3). The average number of co-running operations for 6000 events with Strategy 4 in place for three NN models are 
1.89, 2.04, and 1.74, while without Strategy 4 (but with Strategy 3), the average number is 1.61, 1.62, and 1.52. Hence, Strategy 4 enables a larger number of co-running operations. 

In general, we notice both Strategies 3 and 4 can dynamically change the number of co-running operations, instead of fixing the number of inter-op parallelism as in the traditional TensorFlow (shown as red 
lines in Figure~\ref{fig:timeline}). 

\textbf{Putting all together.}
Figure~\ref{fig:performance}.d shows the performance after applying all strategies together and compares it with the performance of the recommendation and manual optimization. 

We observed that ResNet-50 achieves the largest performance improvement (49\%) among the four NN models. 
Such a large performance improvement largely comes from applying Strategy 3. Many operations in ResNet-50 are not scalable, which brings a lot of opportunities to apply Strategy 3 to co-run operations. Furthermore, ResNet-50 has many small operations which can run together with those time-consuming operations, by applying hyper-threading (Strategy 4). 




\textbf{Comparing with the manual optimization.}
Figure~\ref{fig:performance}.d compares the performance of the manual optimization and our runtime. 
We observed that the performance of ResNet-50, DCGAN and LSTM by our runtime can achieve 8\%, 7\% and 2\% performance improvement than the manual optimization, respectively. 
Our experiments show that for ResNet-50, manual optimization sets intra-op and inter-op parallelisms as 16 and 4. For DCGAN, manual optimization sets them as 34 and 2. For LSTM, manual optimization sets them as 2 and 2.  


For Inception-v3, our runtime performs 2\% worse than the manual optimization. The manual optimization sets intra-op and inter-op parallelisms as 68 and 2, respectively. Such configuration is closing to the configurations chosen by our runtime for most of operations. Hence our runtime performs similar to manual optimization. Our runtime has slight performance loss (2\%). We suspect that the slight performance loss comes from changing intra-op parallelism across operations. 



\section{Discussions}
\label{sec:discussion}


\textbf{Multi-KNL.} Although our work focuses on concurrency control and operation scheduling within a single KNL, our work can work for multiple KNLs. To use multiple KNLs for NN training, the users usually employ either data parallelism or model parallelism~\cite{dean2012large}. 

The data parallelism duplicates the NN model on multiple KNLs, and distributes training data between multiple KNLs. Our runtime system can work on individual KNLs without any change for the data parallelism.

The model parallelism partitions the NN model into multiple groups, each of which is distributed to one KNL. In each KNL, the number of operations available for scheduling is smaller than that in the case of using the single KNL. This indicates that we have less opportunities to co-run operations, but our control over intra-op parallelism should remain the same.

No matter whether the users employ data parallelism or model parallelism, our runtime does not need to be changed.
We leave the evaluation of multiple KNLs as our future work.

\section{Related Work}


\textbf{Performance optimization for dataflow-based frameworks.}
Recent works explore performance optimization for dataflow-based frameworks~\cite{mirhoseini2017device, mirhoseini2018hierarchical, DBLP:journals/corr/abs-1709-02878, gao2018spotlight, 
liu2018pim}.
Mirhoseini et 
al.~\cite{mirhoseini2017device, mirhoseini2018hierarchical} propose a method 
that first schedules the operations to groups and then places those groups onto devices.
Hafner et al.~\cite{DBLP:journals/corr/abs-1709-02878} 
allow the TensorFlow execution engine to parallelize computation to 
improve training performance. 
Liu et al.~\cite{liu2018pim} propose a software and hardware co-design of heterogeneous processing-in-memory system
that schedules NN training operations across compute resources 
to improve hardware utilization.
Our work is different from the existing efforts. We propose runtime scheduling strategies that co-run operations to improve hardware utilization and system throughput on manycore platforms. We also explore performance modeling to predict performance of operations with various intra-op parallelisms, which is not explored in the existing efforts.
\textbf{Thread concurrency throttling.} Previous work explores dynamic thread concurrency throttling to achieve the optimal performance~\cite{di2013regulating, pusukuri2011thread, 
rughetti2012machine}. 
Pusukuri et al.~\cite{pusukuri2011thread} develop a framework to dynamically determine an appropriate number of threads 
that identifies 
near optimal number of threads with OpenMP 
to achieve the optimal performance.
Sanzo et al.~\cite{di2013regulating} proposes a self-regulation approach that 
predicts the scalability of applications to improve performance.

Our concurrency throttling approach differs from them, in that we not only study concurrency for individual operations, but also study inter-op concurrency control by co-running operations 
with various runtime scheduling strategies. 



\section{Preliminary Study on GPU and Future Work}
\label{sec:future_work}

Since using GPU is a common method for NN training, we explore the possibility of employing our method for GPU. In this section, we present our preliminary study on GPU and discuss our future work.


\subsection{Preliminary Study on GPU}

We study the performance of operations on GPU from two perspectives: (1) performance variance with different intra-op parallelisms; (2) performance impact of co-running operations. We use an Nvidia Tesla P100 GPU and CUDA 9. Tesla P100 contains 3584 cores, 56 Streaming Multiprocessors (SMs) and 4MB L2 cache size. TensorFlow uses cuDNN (v7.0 in our study) to execute many operations on GPU. 
cuDNN is not open-source, and we cannot manipulate the intra-op parallelism for those operations. Hence, in our study, we focus on the GPU operations that are not implemented in cuDNN. For each operation, we develop a script to run it standalone as in Section~\ref{sec:background}.
Furthermore, we use input data sizes in the NN model Inception-v3 to study performance of operations.

\textbf{Study of intra-op parallelism.} 
For each operation, we can change either the number of threads per thread block or the number of thread blocks to change the intra-op parallelism.

\begin{figure}[tb!]
	\centering
	\includegraphics[width=0.85\linewidth]{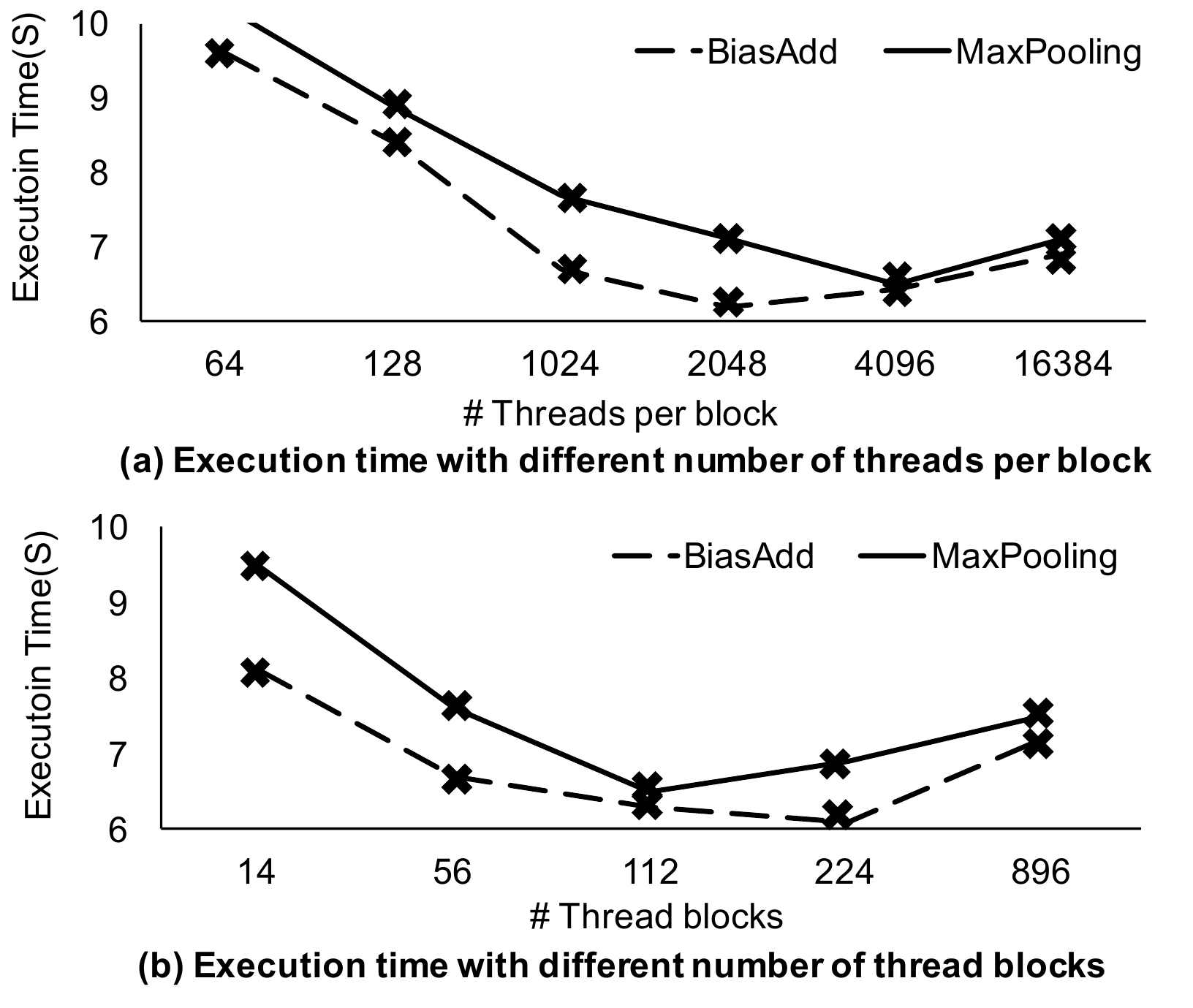}
	\vspace{-5pt}
	\caption{Performance variance of two operations with different intra-op parallelisms on GPU. The reported execution time is the total execution time of ten thousand runs.}
    
    \centering
	\label{fig:gpu_intra}
    \vspace{-5pt}
    
\end{figure}

We first change the number of threads per thread block, while using the TensorFlow's default number of thread blocks (56 in our system). By default, TensorFlow uses 1024 threads per thread block.
We study two common and time-consuming operations, $BiasAdd$ and $MaxPooling$. Figure~\ref{fig:gpu_intra}.a shows the results. The figure reveals that there is a big performance variance across different cases. The case using the default intra-op parallelism in TensorFlow does not result in the best performance. There is up to 18\% performance difference between the default case (i.e., 1024 threads per thread block) and the case with the best performance.

Furthermore, we change the number of thread blocks to run the two operations, while using the TensorFlow's default number of threads per thread bock (1024 in our platform). Figure~\ref{fig:gpu_intra}.b shows the results.
The figure shows there is up to 11\% performance difference between the default case (i.e., 56 thread blocks) and the case with the best performance.

\textbf{Study of inter-op parallelism.} 
To co-run operations in a GPU, we
concurrently run two processes, each of which uses one CUDA stream to run an operation. Note that TensforFlow only supports one CUDA stream to run operations per GPU, hence we cannot simply use the existing CUDA stream in TensorFlow to co-run operations. When co-running operations, we use the optimal intra-op parallelism for each operation. We also use cuDNN to run each operation for best performance.

Table~\ref{tab:gpu_corun} summarizes the results of five operations, $Conv2DBackpropFilter$,  $Conv2DBackpropInput$, $Conv2D$, $BiasAdd$ and $MaxPooling$. These operations take more than 70\% of total training time of ResNet-50, DCGAN and Inception-v3 models.
For each operation, we run two instances of it to enable co-run tests, as in common ML models. The table compares the performance of co-run cases with that of serial cases. Each serial case runs the two instances of an operation in serial. The serial case is the default execution mode in TensorFlow.
Table~\ref{tab:gpu_corun} shows that co-running operations achieves better performance than serial execution in all cases. Co-running operations leads to up to 90\% performance improvement.

\textbf{Conclusions.} Using the default concurrency control in TensorFlow, we cannot achieve the best performance on GPU. This result is consistent with what we observe on CPU. There is a room for us to study on GPU.

\begin{table}
\footnotesize 
\caption{Co-running operations on GPU. The reported execution time is total execution time of ten thousand runs.}
\vspace{-5pt}
\label{tab:gpu_corun}
\centering
\resizebox{0.45\textwidth}{!}{
\begin{tabular}{|c|c|c|c|}
\hline
Operations                            & Strategies           & Time (s) & Speedup \\ \hline
\multirow{2}{*}{Conv2DBackpropFilter} & Serial execution & 9.8      & 1.00    \\ \cline{2-4} 
                                      & Co-run               & 5.5      & 1.78    \\ \hline
\multirow{2}{*}{Conv2DBackpropInput}  & Serial execution & 18.2     & 1.00    \\ \cline{2-4} 
                                      & Co-run               & 9.9      & 1.84    \\ \hline
\multirow{2}{*}{Conv2D}               & Serial execution & 17.4     & 1.00    \\ \cline{2-4} 
                                      & Co-run               & 9.1      & 1.91    \\ \hline
\multirow{2}{*}{BiasAdd}              & Serial execution & 11.8     & 1.00    \\ \cline{2-4} 
                                      & Co-run               & 6.6      & 1.79    \\ \hline
\multirow{2}{*}{MaxPooling}           & Serial execution & 12.6     & 1.00    \\ \cline{2-4} 
                                      & Co-run               & 7.2      & 1.75    \\ \hline
                                      
\end{tabular}
}
\vspace{-15pt}
\end{table}

\subsection{Future Work}


Runtime concurrency control and operations scheduling on GPU faces a couple of challenges: 
First, there is a large search space to decide the optimal intra-op parallelism for an operation. 
Using the hill climbing algorithm, we may need to run a large number of sample cases, which increases performance overhead. 
Second, the cuDNN library is not open-source, which brings a challenge to control intra-op parallelism for each operation.

We plan to address the above challenges as our future work, but we discuss our solutions as follows.

To address the first challenge, we must narrow down the search space. There is a two-dimensional space to control intra-op parallelism on GPU: the number of threads per thread block and the number of thread blocks. This leads to a search space of $\mathcal{O}(n^2)$, where $n$ is the number of configurations in each dimension. We observe that the optimal number of thread blocks seems to be independent of the optimal number of threads per block. This observation allows us to consider the two dimensions independently, and reduces the search space to $\mathcal{O}(2n)$. 
Furthermore, we observe that there is little performance difference between a large number of threads per block and a small one. For example, there is little performance difference (less than 3\% for $BiasAdd$ and $MaxPooling$) between using 10 threads per block and 100 threads per block.
 This allows us to use a rather large interval to further reduce the number of sample cases. 

To address the second challenge, we can replace some time-consuming operations (e.g. convolutions) in cuDNN with operations from an open-source library with better or comparable performance (e.g., ISAAC~\cite{Tillet:2017:IAC:3126908.3126939}) such that we can control intra-op parallelism at runtime.

\section{Conclusions}

\label{sec:conclusion}
The new generation of ML frameworks such as TensorFlow and Caffe2 embraces a dataflow model and represents computation by a directed graph composed of operations. Training an NN model based on such ML frameworks can generate a lot of operations, which brings challenges to manage them for best performance. We expect such challenges will be more pronounced in the future NN models. In this paper, we study how to automatically decide intra-op parallelism for each operation and how to co-run operations to improve performance. We use a performance model-driven approach to guide the runtime system to parallelize and schedule operations. 
Guided by the performance model, we introduce a set of practical and effective scheduling strategies. Applying the performance model and scheduling strategies to the TensorFlow runtime, we achieve great performance improvement. Our work reveals many opportunities to improve the performance of NN training through concurrency control and operation scheduling.
   
\section{acknowledgement}

\label{sec:acknowledgement}
This work was partially supported by U.S. National Science Foundation (CNS-1617967, CCF-1553645 and CCF-1718194) and Chameleon Cloud. This work was also partially supported by the U.S. Department of Energy, Office for Advanced Scientific Computing (ASCR) under Award No. 66150: ``CENATE: The Center for Advanced Technology Evaluation''. Pacific Northwest National Laboratory (PNNL) is a multiprogram national laboratory operated for DOE by Battelle Memorial Institute under Contract DE-AC05-76RL01830.   


\bibliographystyle{IEEEtran}
\begin{spacing}{0.9}
\bibliography{li}
\end{spacing}
\clearpage

\end{document}